\documentclass[%
reprint,
superscriptaddress,
 amsmath,amssymb,
 aps,
prb,
]{revtex4-2}
\usepackage{pdfpages}
\usepackage{pgffor}
\usepackage{graphicx}
\usepackage{dcolumn}
\usepackage{bm}
\usepackage{siunitx}
\setcitestyle{super}

\newcommand*{\citen}{}
\DeclareRobustCommand*{\citen}[1]{%
  \begingroup
    \romannumeral-`\x 
    \setcitestyle{numbers}%
    \cite{#1}%
  \endgroup
}
\usepackage{soul,xcolor}
\soulregister\cite7
\soulregister\ref7
\soulregister\textbf7
\soulregister\textit7
\usepackage{soul,xcolor}
\soulregister\cite7
\soulregister\ref7
\soulregister\textbf7
\soulregister\textit7
\soulregister\textsubscript7

\usepackage[utf8]{inputenc}   
\usepackage[T1]{fontenc}

\makeatletter
\AtBeginDocument{\let\LS@rot\@undefined}
\makeatother

\begin{document}

\preprint{APS/123-QED}
\title{Magnetic field induced modulation of heat transport in CrPS$_4$ probed by TaIrTe$_4$ transverse Seebeck thermometry}



\author{Krishnaraajan Sundararajan$^{\dagger}$}
\thanks{Corresponding author: k.sundararajan@rug.nl}
\affiliation{Zernike Institute for Advanced Materials, University of Groningen, 9747 AG Groningen, The Netherlands}

\author{Sergio Alvarruiz}
\thanks{These authors contributed equally to this work.}
\affiliation{Zernike Institute for Advanced Materials, University of Groningen, 9747 AG Groningen, The Netherlands}

\author{Cédric A. Cordero-Silis}%
\affiliation{Zernike Institute for Advanced Materials, University of Groningen, 9747 AG Groningen, The Netherlands}

\author{Majid Ahmadi}%
\affiliation{Zernike Institute for Advanced Materials, University of Groningen, 9747 AG Groningen, The Netherlands}

\author{Marcos H.D. Guimar{\~a}es}%
\affiliation{Zernike Institute for Advanced Materials, University of Groningen, 9747 AG Groningen, The Netherlands}
\affiliation{Cavendish Laboratory, University of Cambridge, CB3 0US Cambridge, UK} 

\author{Jagoda S{\l}awi{\'n}ska}%
\affiliation{Zernike Institute for Advanced Materials, University of Groningen, 9747 AG Groningen, The Netherlands}

\author{Joseph Barker}%
\affiliation{School of Physics and Astronomy, University of Leeds, Leeds LS2 9JT, UK}

\author{Bart J. van Wees}
\affiliation{Zernike Institute for Advanced Materials, University of Groningen, 9747 AG Groningen, The Netherlands}

\begin{abstract}


Understanding and controlling heat flow in low-dimensional magnetic materials is central to thermoelectrics and spin caloritronics. One of the challenges is the absence of sensitive probes of small, field-dependent modulations in thermal transport. Here we demonstrate transverse Seebeck thermometry using exfoliated TaIrTe$_4$ as a probe of thermal conductivity modulated by an external magnetic field on a van der Waals magnet, CrPS$_4$. With a temperature gradient oriented at $45^\circ$ to its principal crystallographic axes, TaIrTe$_4$ exhibits a transverse thermopower exceeding $200~\mathrm{\mu V\,K^{-1}}$ for 900 nm wide and 30 $\mu$m long electrodes, enabling precise detection of variations in the local temperature gradients. Exploiting this with CrPS$_4$ in a nonlocal geometry, we resolve distinct field dependent thermal conductivity changes associated with both the spin-flip and spin-flop magnetic transitions which we attribute to variations in the magnon–phonon coupling in CrPS$_4$. At 5 K, the transverse Seebeck response is modulated by more than 8$\%$ by an external magnetic field, demonstrating strong magnetic control of heat transport in CrPS$_4$. This all van der Waals platform offers high sensitivity to modulations of thermal conductivity and a broad adaptability to a wide range of exfoliated magnetic systems.


\end{abstract}
\maketitle

\section*{Introduction}

Controlling heat flow at the nanoscale is both a fundamental problem in condensed matter physics and a practical challenge for device engineering \cite{Intro_1, Intro_1b, Intro_2, SpinCalo_1,SpinCalo_2}. Low-dimensional magnets specifically offer a unique platform where lattice, electronic and magnetic degrees of freedom couple strongly and the magnetic excitations, magnons, carry a sizable fraction of the heat current \cite{MagnonThermalTransport,VI3_ThermalHall,CrI3_ThermalHall}. Understanding and actively manipulating the magnetic contribution to thermal conductivity would open pathways to magnetically tunable thermal devices \cite{ThermalValve_1,ThermalValve_2} and help enable efficient heat dissipation, critical for improving performance in electronic and energy applications \cite{Thermal_Management}.\\

Van der Waals antiferromagnets provide an interesting platform for investigating magnetically tunable heat and spin transport, combining nonvolatile magnetic order and ultrafast dynamics enabled by THz-frequency magnons \cite{AFM_1, AFM_2}. Their weak interlayer exchange and low magnetic anisotropy enable efficient control of magnetic order with relatively small fields, strongly modulating susceptibility \cite{SQUID_CPS}, heat capacity and thermal conductivity \cite{Kappa_CC3}, and magnon-mediated spin transport \cite{DKW_1,NL_HighField_CPS}. Since magnons carry both spin and heat, investigating their contribution to heat transport offers a complementary route to understanding the fundamental processes governing magnon mediated spin and heat conduction. \\



Despite significant interest, isolating the role of magnons to thermal conductivity remains challenging with existing probes. Conventional local thermal probes often merge heating and detection pathways, perturb thermal boundary conditions, rely on finite contact areas or transducer layers, and provide limited sensitivity to nonlocal magnon heat transport \cite{Intro_1,Intro_1b,SThM_SameHeaterDetector,RamanThermometry,SuspendedBridge}. Thermal Hall and nonlocal spin Seebeck effect can access magnetic heat transport, but the responses are often weak or intertwined with other transport channels \cite{ThermalHall_1,SpinSeebeck}. Consequently, quantitative studies of magnon thermal transport have remained largely indirect and concentrated on bulk materials or macroscopic films, demonstrating the need for sensitive thermal probes compatible with thin van der Waals magnets. Importantly, since magnon transport experiments are typically performed on exfoliated flakes \cite{DKW_1,NL_HighField_CPS,Mag_MPS,ACS_Nano,luo2026flat,HallBar_CPS,qi2023giant}, thermal transport measurements must be carried out in comparable low-dimensional geometries to ensure a consistent interpretation.\\

These limitations motivate the need for thermal sensors and material platforms that intrinsically convert temperature gradients into an electrical response. Here, we show that goniopolar materials provide a magnetic field free detection scheme based on the transverse Seebeck effect \cite{Uchida_Trans_Seebeck}, enabling sensitive access to thermal transport and in particular to magnetic field induced modulations of the thermal conductivity, which has not been demonstrated previously.\\

Goniopolar materials exhibit axis-dependent carrier polarity, where the dominant transport carrier type (hole-like or electron-like) varies with the crystallographic direction, leading to a positive Seebeck coefficient along one axis and negative along the other \cite{Uchida_Trans_Seebeck}. Such behavior has been reported in bulk single crystals of semi-conductors such as KMgBi \cite{KMgBi} and NaSnAs \cite{NaSnAs} and in (semi-)metallic systems such as NaSn$_2$As$_2$ \cite{he2019fermi}, LaPt$_2$B \cite{LaPt2B}, WSi$_2$ \cite{WSi2} and TaIrTe$_4$ \cite{TaIrTe4_Gonio, TaIrTe4_Gonio_2}. Owing to the opposite signs of the Seebeck coefficients along different crystallographic axes, temperature gradients that donot coincide with principal crystallographic axes, generates a transverse voltage in the absence of a magnetic field, known as the transverse Seebeck effect \cite{Uchida_Trans_Seebeck}. Since the voltage detection direction is decoupled from the thermal gradient direction, the detector sensitivity can be enhanced by increasing the electrode length without increasing the applied temperature gradient.\\


Here we demonstrate the functionality of TaIrTe\textsubscript{4}, a van der Waals two-dimensional semi-metal as a sensitive thermal sensor and utilize it to investigate the magnetic field induced modulation of the thermal conductivity in exfoliated CrPS\textsubscript{4}, a prototypical magnetic insulator. Using a nonlocal geometry where the TaIrTe\textsubscript{4} electrodes are defined at an angle away from the high-symmetry crystallographic axes, we detect a transverse thermoelectric voltage generated by an in-plane temperature gradient via the transverse Seebeck effect. We observe a pronounced field-dependent modulation of the thermal conductivity of CrPS\textsubscript{4} across both the spin-flip and spin-flop transitions. Notably, we observe a clear thermal transport feature associated with the spin-flop transition of CrPS$_4$ which was not detected previously in nonlocal magnon spin transport measurements \cite{ACS_Nano}. We find that the magnetic field induced modulation of the thermal conductivity of CrPS$_4$ cannot be explained by magnetostriction induced changes in the phononic thermal conductivity or by the field evolution of the magnon thermal conductivity alone. Instead, the experimental observations are consistent with a contribution from magnon--phonon coupling, which evolves under an applied magnetic field. These results further show that the proximity of a transverse thermoelectric layer can be utilized as a sensitive, nonlocal and non-invasive probe to investigate the role of magnons in heat transport in an all van der Waals device architecture.


\section*{Results} 

\begin{figure*}[htb]
    \centering
    \includegraphics[width=\textwidth]{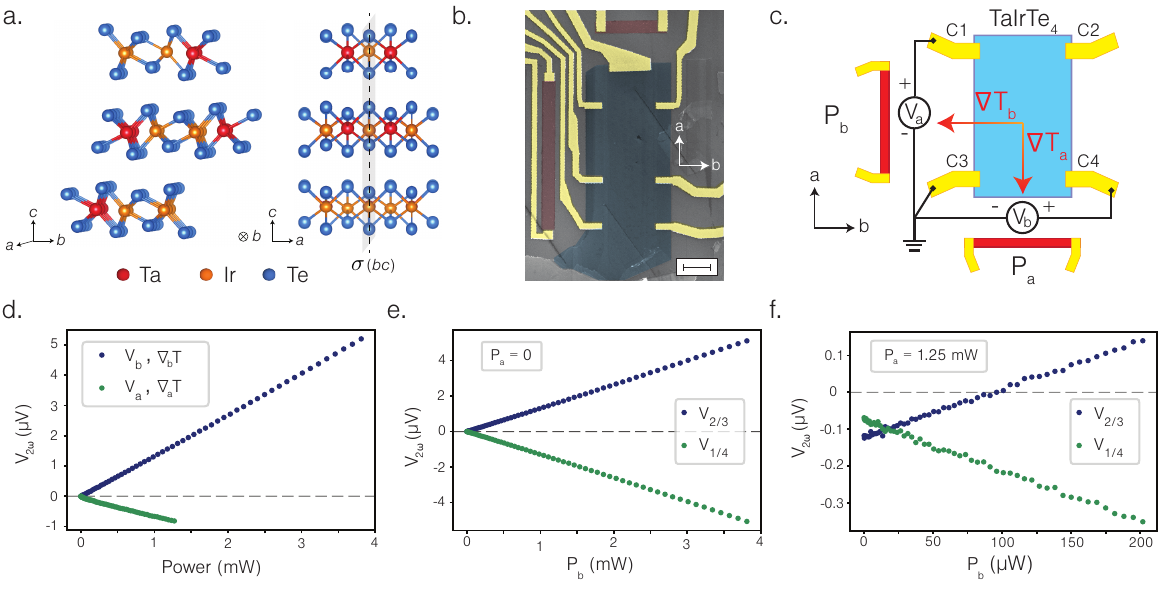}   
    \caption{a) Crystal structure of TaIrTe$_4$, highlighting the mirror symmetry (denoted by \textit{$\sigma(bc)$}), b) false-colored scanning electron micrograph of the device fabricated to verify the goniopolar nature of TaIrTe\textsubscript{4} (scale bar: 5 $\mu$m), c) schematic illustration of the device geometry for probing the transverse Seebeck response: in blue is the TaIrTe\textsubscript{4} flake and in red, the platinum heaters employed, where P\textsubscript{a}(P\textsubscript{b}) refer to the power dissipated in the platinum heaters, generating a temperature gradient along the crystallographic-\textit{a (b)} axes. d) Room temperature longitudinal thermoelectric response (Seebeck) detected across TaIrTe\textsubscript{4} for a temperature gradient along the crystallographic-\textit{a} and -\textit{b} axes (in green and blue respectively). e) The thermoelectric response detected along the diagonals of the flake as a function of the heater power that generates a temperature gradient along the crystallographic-\textit{b} axis and f) the response with an additional temperature gradient along the crystallographic-\textit{a} axis.}
    \label{fig:Gonio}
\end{figure*}

\subsection*{Goniopolar nature and the transverse Seebeck effect in exfoliated TaIrTe$_{4}$ flakes} 

Firstly, we examine the goniopolar character of TaIrTe\textsubscript{4} (Fig. \ref{fig:Gonio}a) in exfoliated thin flakes using two complementary approaches. We fabricated a sample with platinum heaters adjacent to the TaIrTe\textsubscript{4} flake on a Si/SiO$_2$ substrate, where the platinum heaters were aligned to generate a temperature gradient predominantly along the crystallographic-\textit{a} or -\textit{b} axis as shown in Fig. \ref{fig:Gonio}b. An alternating current (I$_{ac}$ $\propto$ $\sin \omega$t) is sent through the Pt heater that generates a temperature gradient through Joule heating. The second-harmonic ($\nabla$T $\propto$ I(t)$^2$ $\propto$ $\sin$ 2$\omega$t) voltage responses are measured across TaIrTe\textsubscript{4} using lock-in amplifiers. Firstly, we examine the goniopolar character by measuring the longitudinal Seebeck coefficient for temperature gradients oriented along the crystallographic-\textit{a} and -\textit{b} directions, similar to that in Ref. \citen{TaIrTe4_Gonio}. A different sign of the Seebeck coefficient along the two crystallographic axes enforces that for temperature gradients along off-axis directions, the generated voltage attains a transverse component in addition to the longitudinal voltage response. Fig. \ref{fig:Gonio}c illustrates the measurement geometry and the convention of the voltage detection used for the thermoelectric measurements. Fig. \ref{fig:Gonio}d shows the longitudinal voltage generated in response to a temperature gradient along the two crystallographic axes, as a function of the heater power at room temperature. Although a quantitative comparison of the effect between the crystallographic-\textit{a} and -\textit{b} axis is not possible due to the asymmetries in the TaIrTe\textsubscript{4} flake and heaters, it can be clearly observed that the voltage responses are opposite in sign depending on whether the temperature gradient is along the crystallographic-\textit{a} or -\textit{b} axis, proving the goniopolar nature of the material. Defining the sign of the Seebeck coefficient according to $E_i=-S_{ii}\nabla_iT$, and using the convention of the measured polarity of the thermoelectric voltages, we find $S_{\text{bb}} > 0$ while $S_{\text{aa}} < 0$, consistent with the previously reported Seebeck coefficients of bulk TaIrTe\textsubscript{4}.\\

To further study this effect, the voltage was simultaneously measured along two different diagonals of the flake. Fig. \ref{fig:Gonio}e shows the voltage response along the diagonals as a function of the heater power that generates a temperature gradient along the crystallographic-\textit{b} axis. With $\nabla$T $\parallel$ $\hat{b}$, we observe that voltage response is approximately equal in magnitude but opposite in sign for both diagonals. When an additional temperature gradient along the crystallographic-\textit{a} axis is simultaneously applied ($\nabla$T $\parallel$ $\hat{a}$), as shown in Fig. \ref{fig:Gonio}f, we observe a shift in the detected voltage response for both the diagonals. We note that for normal metals that do not exhibit a goniopolar nature, such as platinum, the additional temperature gradient would increase the response longitudinal to the temperature gradient (V$_{+/-} \equiv$ V$_{2/3}$) while decreasing the transverse response (V$_{1/4}$). However, for TaIrTe\textsubscript{4}, owing to its goniopolar nature, the additional temperature gradient increases V$_{4/1}$ and reduces V$_{2/3}$, even making the latter zero for an applied power of P$_{\nabla_{a}T}$ = 1.25 mW and P$_{\nabla_{b}T}$= 100 $\mu$W, where P$_{\nabla_{i}T}$ denotes the power dissipated in the heater electrode that generates a temperature gradient oriented along $\nabla_{i}T$. These observations further validate that the goniopolar nature of TaIrTe$_4$ persists even in exfoliated thin flakes ($\sim$ 22 nm thick).
\\

Owing to the opposite signs of the Seebeck coefficients along the crystallographic- 
\textit{a} and -\textit{b} axes, an off-axis temperature gradient induces a measurable transverse voltage in TaIrTe\textsubscript{4}. When the temperature gradient is oriented at an angle $\gamma$ with respect to the crystallographic-\textit{a} axis, the resulting transverse voltage is given by:

\begin{equation}
\label{eq:TSE}
    V_{\text{x}} = -\frac{(S_{aa} - S_{bb})}{2} \, \sin(2\gamma) \, \nabla_{\text{y}} T\,L_{\text{x}},
\end{equation}

where $S_{aa}$ and $S_{bb}$ are the Seebeck coefficients along the crystallographic-\textit{a} and -\textit{b} axes (such that the transverse Seebeck coefficient is given by S$_{xy} = \frac{(S_{aa}-S_{bb}).\sin(2\gamma)}{2}$) and $\nabla_{y} T$ is the applied longitudinal temperature gradient and L$_x$ corresponds to the length of the electrode transverse to the temperature gradient (see Supplementary Section I). We emphasize that the temperature gradient is due to a temperature drop along $\hat{y}$ while the corresponding length with which the voltage scales is along an $\hat{x}$. This allows the sensitivity of the thermal sensor to be enhanced by increasing the electrode length, as the generated signal scales proportionally with the the length perpendicular to the temperature gradient, making the transverse thermopower demonstrated here in TaIrTe\textsubscript{4} particularly advantageous for nonlocal thermal measurements.


\subsection*{Effect of magnons on heat transport in CrPS$_{4}$}

\begin{figure*}[htb]
    \centering
    \includegraphics[width=\linewidth]{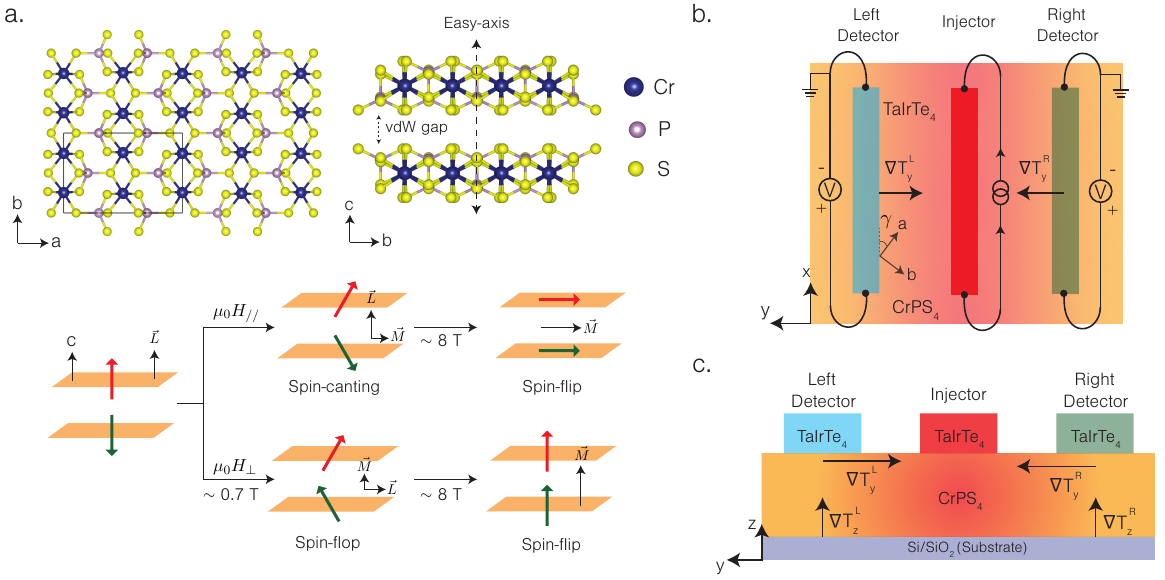}   
    \caption{a) (Top panel) Crystal structure of CrPS$_4$ in the \textit{ab} and in the \textit{bc} planes. (Bottom panel) Illustration of the magnetic transitions in CrPS$_4$ under an external magnetic field with the corresponding directions of the N\'eel ($\vec{L}$ = $\vec{m_1}$ - $\vec{m_2}$) and the magnetization ($\vec{M}$ = $\vec{m_1}$ + $\vec{m_2}$) vectors, where $\vec{m_1}$ and $\vec{m_2}$ denote the sub-lattice magnetization. Illustration of the nonlocal geometry along with the various temperature gradients viewed along (b) the xy plane and (c) the yz plane.}
    \label{fig:Geometry}
\end{figure*}

Next, we utilize the transverse Seebeck effect in TaIrTe\textsubscript{4} to investigate the role of magnons in heat transport in an insulating van der Waals magnet. We choose CrPS\textsubscript{4} due to its well-established magnetic properties, including magnon band structure resolved by inelastic neutron scattering \cite{Calder_PhysRevB_102_024408_2020}, magnetization characterized by SQUID magnetometry \cite{SQUID_CPS,bud2021magnetic}, and study of nonlocal magnon spin transport \cite{DKW_1,NL_HighField_CPS,ACS_Nano, arxiv_Aniso} and thermally driven magnons via the (non)local spin Seebeck effect \cite{DKW_1,NL_HighField_CPS,HallBar_CPS,qi2023giant,arxiv_Aniso}. CrPS\textsubscript{4} is a layered van der Waals A‑type antiferromagnetic insulator, wherein Cr$^{3+}$ ions couple ferromagnetically within each layer. Owing to its weak interlayer antiferromagnetic exchange coupling and out-of-plane easy axis anisotropy, CrPS\textsubscript{4} exhibits a spin-flop transition at approximately 0.7~T under an out-of-plane magnetic field, before reaching a fully saturated collinear state around 8~T. In contrast, for an in-plane magnetic field, perpendicular to the easy axis, the spins cant continuously towards the field and eventually saturate into a collinear state at approximately 8~T, as illustrated in Fig. \ref{fig:Geometry}a. Across the spin-flip transition, the net magnetization ($\vec{M}$) saturates while the N\'eel vector ($\vec{L}$) collapses to zero.
\\

To investigate magnon-mediated heat transport in CrPS\textsubscript{4} ($\sim$30--60~nm), we employ a nonlocal geometry as shown in Fig.~\ref{fig:Geometry}b. Heat is injected and detected without driving a charge current through CrPS\textsubscript{4}, avoiding magnetoresistive contributions. This geometry furthermore suppresses the temperature gradient along the detector, $\nabla_xT$, thereby enhancing the transverse thermoelectric response generated by $\nabla_yT$. Moreover, the nonlocal thermopower voltage described by Eq.~\ref{eq:TSE} allows the measurement of the temperature gradient across the TaIrTe\textsubscript{4} detector. A magnetic field induced increase (decrease) in the thermal conductivity of CrPS\textsubscript{4} therefore reduces (enhances) this gradient and produces a corresponding modulation of the measured thermoelectric voltage.\\


\begin{figure*}[htb]
    \centering
    \includegraphics[width=\textwidth]{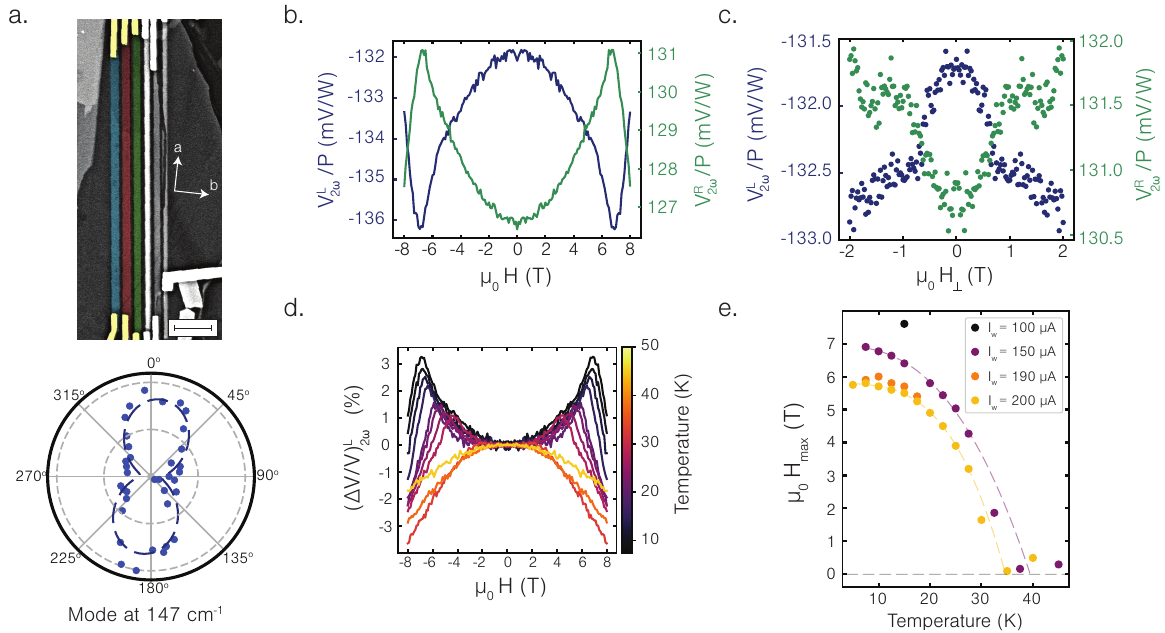}   
    \caption{a) (Top) Scanning electron micrograph (false colored: red indicating the heater electrode, blue and green indicating the detectors on the left and the right side, scale bar: 4 $\mu$m) of device D1 and (bottom) Raman intensities of the mode corresponding to 147 cm$^{-1}$ a function of the sample rotation angle. The solid dots correspond to the experimental data and the dotted lines to the fitting curve; Field dependence of the symmetrized second-harmonic voltage response measured by the left and the right detectors b) across the spin-flip transition of CrPS$_{4}$ for an applied in-plane magnetic field and c) across the spin-flop transition for an applied out-of-plane magnetic field (at 7.5K and I$_{heater}$ = 150 $\mu$A). d) The normalized second-harmonic voltage response across TaIrTe$_4$ as a function of the applied external in-plane magnetic field for varying temperature. e) Temperature dependence of the field corresponding to the maximum in the second-harmonic voltage response ($\sim$ spin-flip field) modulation for varying heater currents. }
    \label{fig:D1}
\end{figure*}

We fabricated four devices in a nonlocal geometry to investigate the magnetic field dependence of the thermal conductivity of CrPS$_4$. The first device (D1) in Fig. \ref{fig:D1}a (top), has TaIrTe\textsubscript{4} electrodes defined at an angle of $\gamma$ $\sim$ 8.8$^{o}$ relative to the crystallographic-\textit{a} axis, as confirmed by the polarized Raman measurements \cite{Liu2018_raman, Liu2023_raman} shown in  Fig. \ref{fig:D1}a (bottom) (see Supplementary Section II). Similar to the measurement protocol explained before, the temperature gradient in these devices are generated by sending an alternating current (I$_{ac}$ $\propto$ $\sin \omega$t) through TaIrTe\textsubscript{4}, that generates a temperature gradient normal to the strip, corresponding to $\hat{y}$ in Fig. \ref{fig:Geometry}b. 
\\

The second-harmonic voltage responses are measured simultaneously at detectors located on either sides of the heater, allowing for decoupling the effects arising from in-plane temperature gradients (along $\hat{y}$) from an out-of-plane temperature gradient (along $\hat{z}$), as illustrated in Fig. \ref{fig:Geometry}c. We use blue and green to indicate the second-harmonic voltage response measured across the left (V$^{L}_{2\omega}$) and the right detectors (V$^{R}_{2\omega}$) respectively. Fig. \ref{fig:D1}b shows the second-harmonic voltage response as a function of the applied magnetic field at 7.5 K. In the absence of a magnetic field, we observe that the voltages measured across the left and right detectors are comparable in magnitude but opposite in sign, indicating that the response originates from an in-plane transversal temperature gradient (namely $\nabla_y$T). We further verify that the detector sensitivity originates from the transverse Seebeck effect using control device D0, wherein TaIrTe\textsubscript{4} is aligned along a high-symmetry axis. The absence of opposite voltage polarities at the two detectors supports that a finite off-axis alignment, $\gamma \neq 0$, is required to generate the transverse thermoelectric response (see Supplemetary Section V).\\



Although the zero-field transverse thermopower reflects the temperature dependence of the thermal conductivity of CrPS\textsubscript{4}, it is not determined by this contribution alone. It also contains the temperature dependent contributions from the interfacial Kapitza resistance, the Seebeck coefficients of TaIrTe\textsubscript{4}, and the thermal conductivity of the Si/SiO$_2$ substrate preventing a reliable estimation of the temperature dependence of the thermal conductivity of CrPS$_4$. (see Supplementary Section VI). Furthermore, owing to the nonlocal geometry employed here, it is not possible to directly measure $\nabla_yT$ that generates the transverse thermopower. As a result, the ordinary Nernst effect (ONE) is used to quantitatively extract the thermoelectric coefficients (see Supplementary Section V) and we estimate that the transverse Seebeck coefficient reaches a value of S$_{xy} \sim$ 70 $\mu$V/K for $\gamma$ $\sim$ 8.8$^{\circ}$. We note that owing to the nonlocal geometry employed here, independent estimation of S$_{aa}$ and S$_{bb}$ is not possible.\\


Apart from its temperature dependence, the thermal conductivity of CrPS\textsubscript{4} is strongly modified by an applied magnetic field. We note that the detector sensitivity is enabled by the transverse Seebeck effect of TaIrTe$_4$, $S_{xy}$, whereas the field-dependent modulation reflects changes in the longitudinal thermal conductivity of CrPS$_4$, namely $\kappa_{yy}$. By Onsager reciprocity relations \cite{Onsager_A,Application_Onsager,Buttiker_TenFold}, the longitudinal thermal conductivity is even under magnetic field reversal, $\kappa_{yy}(B)=\kappa_{yy}(-B)$ and we therefore analyze the thermoelectric response symmetrized with respect to the applied magnetic field.\\

We exclude possible contributions from the intrinsic magnetothermopower of TaIrTe$_4$, its magnetoresistance, thermally driven magnon spin transport and the Righi–Leduc effect as the origin of the observed field-dependent modulation (see Supplementary Section IV) and attribute the modulation primarily to magnetic field induced changes in the thermal conductivity of CrPS$_4$. We further validate this interpretation by employing the well-established $3\omega$ harmonic detection technique \cite{Cahill_3omega,Tutorial_3omega} to measure magnetic field induced modulations of the thermal conductivity of CrPS\textsubscript{4} using a platinum electrode (see Supplementary Section VII).\\

\begin{figure*}[htb]
    \centering
    \includegraphics[width=\textwidth]{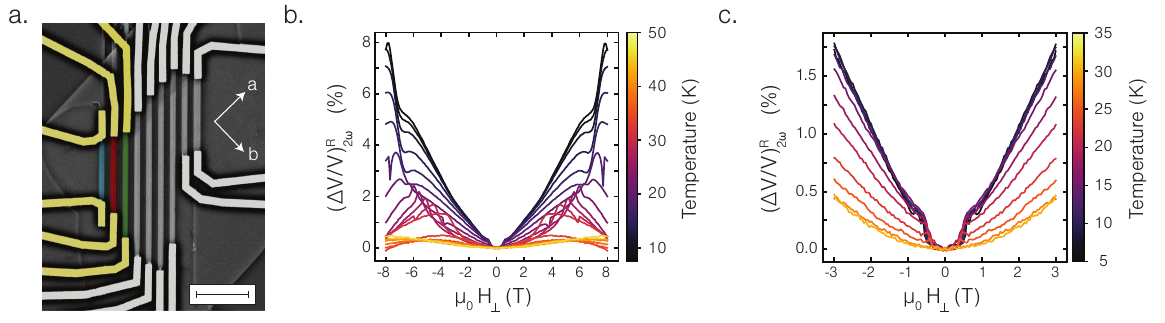}   
    \caption{a) False colored scanning electron micrograph of device D2; Field dependence of the symmetrized and normalized second-harmonic voltage response of the right detector b) for an out-of-plane magnetic field across the spin-flip transition and c) an out-of-plane magnetic field across the spin-flop transition of CrPS$_4$ as a function of temperature (I$_{heater}$ = 30 $\mu$A). (Scale Bar: 5 $\mu$m) }
    \label{fig:D2}
\end{figure*}

Fig.~\ref{fig:D1}b shows the symmetrized second-harmonic voltage response of the TaIrTe\textsubscript{4} detectors as a function of applied magnetic field. We normalize the second-harmonic voltage response to its zero-field value and define $\Delta V/V=[V(B)-V(0)]/V(0)$. Since the thermoelectric voltage response is proportional to the local temperature gradient, $\Delta V/V$ directly reflects the relative magnetic field induced modulation of the temperature gradient across the detector. Furthermore, this normalization is insensitive to the direction of the in-plane temperature gradient, since both $V(B)-V(0)$ and $V(0)$ reverse sign upon reversal of the thermal gradient (see Supplementary Section III). For fixed heating power, a positive $\Delta V/V$ therefore corresponds to an enhanced thermoelectric response and hence a decrease in the thermal conductivity. We observe the second-harmonic voltage response increases approximately quadratically up to the spin-flip field of CrPS\textsubscript{4} followed by a sharp decrease at higher fields. We observe that the modulation is qualitatively isotropic, exhibiting a similar field dependence irrespective of the direction of the applied in-plane magnetic field (see Supplementary Section V).\\


Under an out-of-plane magnetic field, we observe a field-dependence of the second-harmonic voltage response corresponding to the spin-flop transition of CrPS\textsubscript{4} as shown in Fig. \ref{fig:D1}c. While the spin-flop transition is visible only at the lowest temperatures and heater currents for device D1, we observe the spin-flip transition to persist over a broader temperature range. Fig. \ref{fig:D1}d shows the second-harmonic voltage response as a function of applied magnetic field for different temperatures. We observe that the feature associated with the spin-flip transition shifts to lower magnetic fields and becomes progressively weaker with increasing temperature. The corresponding peak position, shown in Fig.~\ref{fig:D1}e, follows qualitatively the same temperature-dependent trend as the spin-flip field of CrPS\textsubscript{4} reported previously \cite{bud2021magnetic}. \\

To further enhance the sensitivity of the thermal sensor, an additional device (D2) (Fig. \ref{fig:D2}a) was fabricated wherein the TaIrTe\textsubscript{4} electrodes were defined with a larger offset angle $\gamma$, namely $\gamma \sim 45^{\circ}$. The second-harmonic voltage response as a function of out-of-plane magnetic field is shown in Fig.\ref{fig:D2}b. Owing to the improved sensitivity of the detector strip, S$_{xy}$, these measurements can be performed at substantially lower heater current minimizing the local heating of CrPS$_4$. As seen in Fig. \ref{fig:D2}c, we observe a more pronounced response of the second-harmonic voltage response associated with the spin-flop transition of CrPS\textsubscript{4} indicating a slower decrease in the thermal conductivity of CrPS\textsubscript{4} across the spin-flop transition. In addition, a distinct regime emerges as a function of temperature, namely that for temperatures below 20K, before the spin-flip transition, we observe an increase in the thermal conductivity followed by a sharp decrease until the spin-flip field of CrPS\textsubscript{4}, which is reproducible for an applied in-plane magnetic field and across different samples (see Supplementary Section V). We emphasize that the transverse Seebeck thermometry introduced here provides substantially higher sensitivity to field-induced changes in the thermal conductivity of CrPS$_4$ than the previously reported thermal-bridge measurements \cite{shao2025abnormal}, enabling us to clearly resolve even the spin-flop transition.\\

We note that the modulation of the thermal conductivity of CrPS$_4$ as a function of magnetic field is also reflected on all the other responses that depend either on the temperature rise ($\Delta T$(B)) or on the temperature gradient ($\nabla T$(B)). The influence of the modulation of the thermal conductivity of CrPS\textsubscript{4} on the local first-harmonic voltage response of the heater electrode and on the ONE response measured of the detector are discussed in detail in the Supplementary Section IX, while higher-order responses \cite{nakagawa} are discussed in  Supplementary Section VIII. In our previous work that utilizes a WTe\textsubscript{2}/CrPS\textsubscript{4} heterostructure \cite{ACS_Nano}, we observed an additional feature superimposed on the ONE response measured across WTe\textsubscript{2}, the origin of which was not fully understood. The magnetic field-dependent modulation of the thermal conductivity of CrPS$_4$ established here offers a consistent framework for interpreting this feature. 


\section*{Discussion} 

We briefly summarize the key experimental observations. In the paramagnetic phase above the N\'eel temperature of CrPS$_4$, the thermal conductivity of CrPS$_4$ increases with applied magnetic field, as shown in Fig.~\ref{fig:D1}d. In the magnetically ordered phase, the field dependence becomes non-monotonic. As seen in Fig.~\ref{fig:D1}d, the thermal conductivity decreases on approaching the spin-flip transition and increases sharply beyond the spin-flip field. In Fig.~\ref{fig:D2}b, with lower-heater currents, we observe that the thermal conductivity initially increases before decreasing sharply towards the spin-flip transition. For an out-of-plane magnetic field, the thermal conductivity continues to decrease across the spin-flop transition, although the rate of decrease is reduced across the spin-flop transition.\\

To understand the origin of the field-dependent modulation of the thermal conductivity in CrPS$_4$, we first note that the total thermal conductivity comprises of electronic, phononic, and magnonic contributions. Since CrPS$_4$ is an electrical insulator, the electronic contribution is expected to be negligible, leaving phonons and magnons as the dominant heat carriers. We first consider whether the observed field dependence of the thermal conductivity could originate from a purely phononic mechanism mediated by magnetostriction. An applied magnetic field can modify the lattice parameters through magnetoelastic coupling, thereby altering the phonon spectrum and consequently the phononic contribution to the thermal conductivity. Within a Landau free-energy description, the lowest-order magnetoelastic coupling in an antiferromagnet can be written in the general form
\begin{equation}
F_{\mathrm{me}}=\sum_{ijkl}\epsilon_{ij}\left(C^{MM}_{ijkl}M_kM_l+C^{LL}_{ijkl}L_kL_l + C^{ML}_{ijkl}M_kL_l\right),
\end{equation}
where F$_{me}$ denotes the magnetoelastic free energy, $\epsilon_{ij}$ denotes the strain tensor, $\vec{M}$ is the net magnetization, and $\vec{L}$ is the antiferromagnetic N\'eel order parameter. The tensors $C^{MM}_{ijkl}$, $C^{LL}_{ijkl}$, and $C^{ML}_{ijkl}$ describe the symmetry-allowed magnetoelastic coupling between the lattice strain and the magnetic order parameters. Consequently, a magnetostriction induced modification of the phonon dispersion, and hence of the phononic contribution to the thermal conductivity, would be expected to follow the field evolution of these order parameters. As shown in Fig. \ref{fig:D1}d and Fig. \ref{fig:D2}b, the experimentally observed thermal conductivity modulation beyond the spin-flip transition does not follow the corresponding evolution of the magnetic order parameters, wherein the magnitude of $\vec{M}$ saturates while $\vec{L}$ collapses to zero as shown in Fig.  \ref{fig:Geometry}a. This discrepancy disfavors magnetostriction as the origin of the observed field dependence of the thermal conductivity beyond the spin-flip transition in CrPS$_4$. Possible piezo-magnetic contributions introduce additional symmetry-allowed terms of the form $\epsilon_{ij}H_kL_l$ or $\epsilon_{ij}H_kM_l$. However, these remain governed by the field evolution of the magnetic order parameters and are therefore likewise inconsistent with the observed field dependence of the thermal conductivity.\\

\begin{figure*}[htb]
    \centering
    \includegraphics[width=\textwidth]{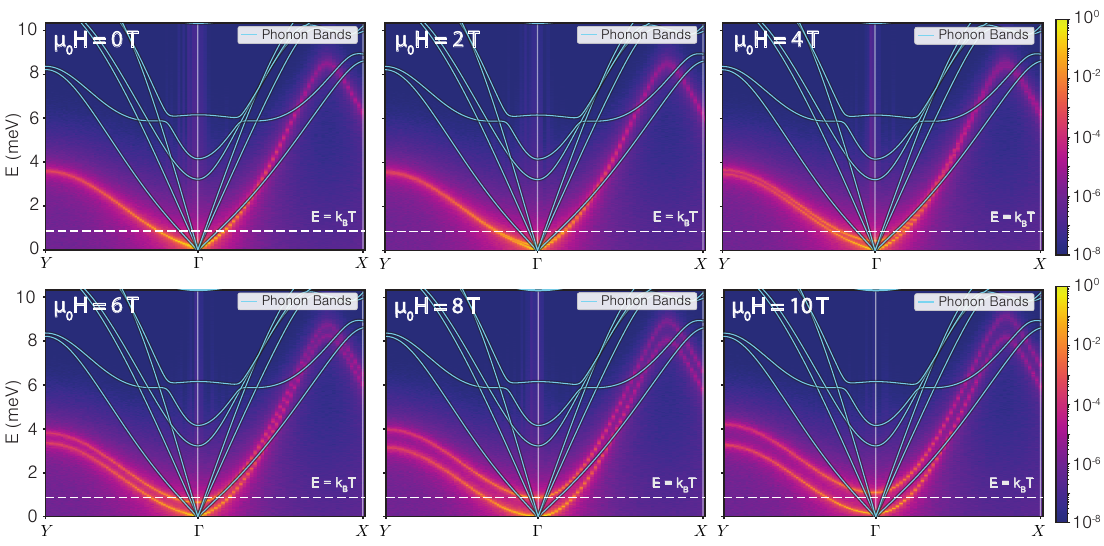}   
    \caption{Magnon and phonon spectrum for increasing magnitude of in-plane magnetic field $\mu_0\vec{H} \cdot\hat{a}$ calculated using atomistic spin dynamics (at $T=10K$) and density functional theory respectively. The color bar corresponds to the dimensionless basis summed magnon power S$^{+-}$ which is proportional to the magnon number density.}
    \label{fig:bands}
\end{figure*}

To understand the magnonic contribution to heat transport, we first calculate the evolution of the magnon band structure under an applied magnetic field using atomistic spin dynamics based on a Heisenberg model parametrized from inelastic neutron-scattering measurements of CrPS$_4$ (see Methods for details). The calculated magnon band structure as a function of the applied in-plane magnetic field along the crystallographic-\textit{a} axis with the corresponding phonon band structure is shown in Fig. \ref{fig:bands}. We note that the uncertainty in the present analysis is the magnon lifetime in CrPS$_4$. In the absence of an experimentally established value, the Gilbert damping parameter used in the calculations are chosen to be $\alpha_{G} = 10^{-3}$. Furthermore, there may be systematic changes in the magnon lifetime across the magnon spectrum that also evolve with an applied magnetic field~\cite{Regout_2605_07416_2026}.\\


Within the magnetically ordered phase, the low N\'eel temperature and the small magnon gap of CrPS$_4$ allows for a substantial population of low-energy magnetic excitations that can be easily modulated by an applied magnetic field. This allows for the possibility of a non-negligible magnon contribution to the thermal conductivity. As shown in Fig. \ref{fig:bands}, at zero applied magnetic field, the two magnon branches are non-degenerate due to magnetic dipole interactions and exhibit lower group velocities than the acoustic phonon branches. However, due to the difference in the dispersion relations themselves, which give rise to a larger magnon density of states at low energies, could potentially produce a sizeable magnetic contribution to the heat capacity. The magnitude of the magnon thermal conductivity is governed by the combined contributions of the magnon heat capacity, group velocity, and lifetime. Since the magnon lifetime in CrPS$_4$ is presently not known, a quantitative comparison between the magnon and phonon contributions cannot be made. Nevertheless, the large density of low-energy magnetic excitations suggests that magnon-mediated heat transport may constitute an appreciable contribution to the thermal conductivity of CrPS$_4$ at low temperatures.\\

We note that applying a magnetic field substantially modifies the magnon spectrum and consequently the magnon density of states (DOS). We observe from Fig.~\ref{fig:bands} two competing effects. Firstly, the low-energy dispersion near the $\Gamma$ point evolves from an approximately linear in the antiferromagnetic phase towards a quadratic ferromagnetic-like dispersion upon approaching the spin-flip transition around \(8\mathrm{T}\). Furthermore, the magnon gap softens and closes at the spin-flip transition. These combined enhance the low-energy thermal magnon population, consistent with the SI Fig.~S22. Secondly, the applied magnetic field causes one of the magnon branches to shift progressively towards higher energies with increasing field, which suppresses its thermal occupation, which is partially compensated by the accompanying change in the DOS as shown in SI Fig.~S22. The field dependence of the thermal magnon population therefore reflects the combined effects of the changing magnon gap, dispersion curvature, and field induced splitting of the magnon branches.\\

These competing field induced effects in the magnon spectrum make it difficult to infer the evolution of the magnon thermal conductivity from the band structure alone. We therefore calculate the magnon thermal conductivity directly from the atomistic spin-dynamics simulations using the Kubo formalism (see Supplementary Section XIII). We find that the magnon thermal conductivity in all directions initially decreases with increasing field ($\mu_0 H = 0-6$~T, as seen in Fig.~S24). The magnitude of this suppression is however anisotropic, reflecting the anisotropic magnon dispersion along the different directions.  We find that the lower the magnon group velocity, the stronger the relative suppression of the magnon contribution to the thermal conductivity. We find that the suppression is approximately $\kappa_{xx}\sim60\%$, $\kappa_{yy}\sim35\%$ and $\kappa_{zz}\sim80\%$ along the different directions. Approaching the spin-flip field, the magnon thermal conductivity begins to increase again as the lower magnon branch acquires a more quadratic dispersion alongside the closure of the magnon gap. These changes enhance the contribution of low-energy magnons and appear to compensate for the reduction in thermal occupation associated with the upper magnon branch shifting to higher energies. As a result, the magnon thermal conductivity recovers towards, and in the case of \(\kappa_{zz}\) even exceeds, its zero-field value. A pronounced sharp increase in the thermal conductivity is observed at the spin-flip transition, followed by a broader enhancement extending to approximately \(10~\mathrm{T}\) (see SI Fig. S24).\\

If the observed field dependence of the thermal conductivity of CrPS$_4$ were governed solely by the magnon contribution, the nonlocal second-harmonic voltage response would be expected to first increase, then decrease on approaching the spin-flip transition, and increase again above it. This characteristic field dependence is not observed experimentally, as shown in Figs.~\ref{fig:D1}d and \ref{fig:D2}b. A similar discrepancy is found across the spin-flop regime. Our spin-dynamics calculations show essentially no structure in the magnon thermal conductivity for an out-of-plane magnetic field between \(0\)--\(1~\mathrm{T}\), with only a weak variation in magnitude for all transport directions (see SI Fig.~S25). In contrast, the experiment shows a clear field dependence across the spin-flop transition as seen in Fig. \ref{fig:D2}c. The opposite sign of the field dependence approaching the spin-flip transition and above it combined with the experimentally observed modulation across the spin-flop transition indicate that modulation of the magnon thermal conductivity alone cannot account for the magnetic field dependence of the total thermal conductivity of CrPS$_4$. \\


This discrepancy points to an additional mechanism, such as magnon--phonon coupling, playing an important role in the magnetic field induced modulation of heat transport in CrPS$_4$. In particular, our spin-dynamics calculations yield \(\kappa_{bb}^{\mathrm{mag}}>\kappa_{aa}^{\mathrm{mag}}\), whereas experimentally measured thermal conductivity reported in Ref.~\citen{arxiv_Aniso} exhibits the opposite anisotropy, with \(\kappa_{bb}<\kappa_{aa}\) at \(25~\mathrm{K}\) and zero applied magnetic field. In a regime where heat transport is predominantly phononic, magnetic excitations can reduce, rather than enhance, the total thermal conductivity by providing an additional scattering channel for phonons. Such magnon--phonon scattering has been shown to strongly suppress lattice thermal transport in a range of magnetic insulators \cite{Mag_Ph1,Kappa_CC3,Mag_Ph2,hentrich2018unusual}.\\






We therefore focus on the phonon contribution to the thermal conductivity and investigate the role of magnon--phonon coupling. Within the Debye--Callaway model \cite{Callaway1,Callaway2}, phonons carry heat while their propagation is limited by scattering processes. Magnon--phonon coupling provides an additional scattering channel and a stronger coupling increases phonon scattering, shortening the phonon mean free path, and therefore suppresses the thermal conductivity. Conversely, a reduction in magnon--phonon scattering allows phonons to propagate more efficiently, leading to an enhanced thermal conductivity.\\

In a magnetic insulator, an applied magnetic field can enhance the thermal conductivity in the paramagnetic phase by suppressing spin fluctuations and thereby reducing spin--phonon scattering, as observed in bulk CrCl$_3$ \cite{Kappa_CC3}. Although coherent magnons are absent in paramagnetic CrPS$_4$, fluctuating spins can still scatter phonons, which is expected to weaken as the magnetization increases. This would enhance the thermal conductivity due to reduced phonon scattering and consequently, for $T > T_N$, we expect a decrease in the second-harmonic voltage response across the TaIrTe$_4$ strip, with increasing magnetic field, which is consistent with our experimental observations in Fig. \ref{fig:D1}d. We note that this trend is less pronounced in Fig. \ref{fig:D2}b, potentially due to an additional background contribution from the intrinsic magneto-Seebeck response of TaIrTe$_4$ for an out-of-plane applied magnetic field. By contrast, for an in-plane magnetic field, where this background contribution is expected to be reduced, the field induced increase in thermal conductivity above the N\'eel temperature is clearly observed, as shown in SI Fig.~S5. \\

Below the N\'eel temperature, assuming phonons dominate heat transport, an applied magnetic field modifies the magnonic band structure, thereby altering the strength of the magnon--phonon coupling, resulting in a modulation of the heat conductivity. To further understand the magnon--phonon interaction in CrPS$_4$, we compare the magnon dispersion obtained from atomistic spin-dynamics calculations with the phonon dispersion calculated using density functional theory (see Methods for details). Fig. \ref{fig:bands} shows the evolution of the magnon spectra with applied magnetic field alongside the corresponding phonon dispersion. Along $\Gamma-X$, we observe that the magnon branch overlaps with the phonon spectrum over an extended range of wavevectors and energies while along $\Gamma-Y$, the magnon and phonon branches predominantly intersect at discrete wavevectors, resulting in a more restricted spectral overlap. The spectral overlap of the magnon and phonon branches identify regions in energy--momentum space where magnon--phonon interactions may become particularly relevant. We emphasize, however, that spectral overlap provides only a qualitative indication of where such interactions may occur and does not directly determine the magnon--phonon scattering rate or hybridization strength. The magnitude of the interaction is determined by the corresponding magnon--phonon coupling matrix elements, which depend on the microscopic spin--lattice coupling, mode polarization, and crystal symmetry \cite{jensen1975spin,cui2023chirality,fang2025efficient}.\\



Since the magnon spectrum evolves continuously with magnetic field, the spectral overlap for magnon--phonon interactions are likewise expected to be field dependent. Along $\Gamma-X$, increasing the field shifts the relevant magnon branches away from the phonon modes, progressively reducing their spectral overlap. Furthermore, above the spin-flip transition, the increasing magnon excitation energies further suppresses their thermal population. If magnon--phonon scattering limits the heat transport, the combined reduction in spectral overlap and magnon population should weaken this scattering channel, allowing phonons to propagate more effectively and would eventually drive the system towards a predominant phononic heat transport at high magnetic fields. This provides a possible microscopic explanation for the experimentally observed increase in thermal conductivity above the spin-flip transition. Additionally, magnon--phonon interactions may play an important role, particularly in the range of $6-8$~T and may contribute to the complex field dependence observed experimentally in Fig.~\ref{fig:D2}b.\\


In the spin Seebeck effect, spectral overlap between magnon and phonon branches has been shown to strongly enhance spin transport through the ensuing formation of magnon--polarons that allows the magnons to inherit the comparatively longer lifetime of the phononic mode~\cite{Kikkawa_PhysRevLett_117_207203_2016,Ramos_NatCommun_10_5162_2019}. In the case of heat transport, magnon--phonon hybridization can reduce the total thermal conductivity since both magnons and phonons carry heat, while hybridization mixes their transport properties such that the shorter-lived component can limit the lifetime of the resulting hybrid modes~\cite{Flebus_PhysRevB_95_144420_2017}. In CrPS$_4$, we note that the enhanced magnon and phonon spectral overlap close to the spin--flip field is coincidental, making it hard to disentangle the effects of hybridization from that of the spin--flip transition on the thermal conductivity. Furthermore, the pronounced anisotropy in the spectral overlap and crossings between the magnon and phonon branches along different crystallographic directions suggests that the conditions for magnon--phonon interactions are strongly anisotropic as well. This may consequently give rise to a corresponding anisotropy in the magnetic field induced modulations of heat transport in CrPS$_4$, in particular between $\kappa_{aa}(B)$ and $\kappa_{bb}(B)$.\\

We emphasize that the primary objective of the present work is to establish TaIrTe$_4$, through its transverse Seebeck response, as a platform for probing magnetic field induced modulations of the thermal conductivity of CrPS$_4$ and establish distinct field-dependent modulations in the heat transport across both the spin-flip and spin-flop transitions of CrPS$_4$. We do not focus on the intrinsic anisotropy of thermal transport in CrPS$_4$, including the difference between $\kappa_{aa}(B)$ and $\kappa_{bb}(B)$ or the dependence on the relative orientation of the magnetic field and heat flow. The measurement platform introduced here, provides a natural route for addressing these anisotropies in future work.\\




\section*{Conclusion}

In conclusion, we demonstrate that the crystallographic axis orientation of TaIrTe$_4$, owing to it's goniopolar nature provides an effective control parameter for tuning its thermoelectric response. In particular, off-axis alignment enhances the transverse Seebeck effect, enabling TaIrTe$_4$ to serve as a sensitive thermocouple for probing magnetic field dependence of heat transport in insulating magnets in a nonlocal geometry. Using this approach, we establish an all van der Waals platform and utilize it to study the field induced modulations of the heat conductivity of the layered antiferromagnet CrPS$_4$, wherein we directly resolve modulations in the thermal conductivity across both the spin-flop and spin-flip transitions of CrPS$_4$, which we attribute to magnetic field induced changes in magnon--phonon coupling. Notably, while signatures of the spin-flop transition have not been observed in nonlocal magnon spin transport measurements in CrPS$_4$ \cite{ACS_Nano}, our heat transport measurements reveal clear and reproducible features. By probing CrPS$_4$ flakes with thicknesses comparable to those used in nonlocal spin Seebeck experiments, our measurements provide a better understanding of heat transport which could potentially help understanding the nonlocal spin Seebeck studies. More broadly, the transverse Seebeck response of TaIrTe$_4$ provides a simple and versatile means of probing heat transport in van der Waals magnetic insulators within a fully two-dimensional device geometry. This platform opens a route to systematic studies of magnetic field dependent thermal transport and magnon-mediated heat conduction in low-dimensional magnetic systems such as CrCl$_3$, VI$_3$, CrOCl, and MnPS$_3$.

\clearpage

\section*{Methods}

\subsection*{Device fabrication} 

Bulk TaIrTe$_{4}$ and CrPS$_{4}$ were purchased from HQ Graphene. CrPS$_{4}$ flakes were exfoliated onto a Si/SiO$_{2}$ (285 nm) substrate in a glovebox under inert conditions (N$_{2}$ glovebox, $<$0.1 ppm O$_{2}$, $<$0.5 ppm H$_{2}$O) using standard scotch-tape method. The measured devices were fabricated by a dry transfer technique \cite{1_Zomer} of a TaIrTe$_{4}$ flake (selected by means of optical contrast) onto the CrPS$_{4}$ flake under inert conditions. Standard electron beam lithography with PMMA as the resist was used to define the heaters for the sample used to verify the goniopolar nature of TaIrTe$_{4}$. The heater electrode used is ($\sim$ 40 nm thick) platinum that was deposited by DC sputtering. Before the deposition of the Ti/Au (5/90 nm) contacts, TaIrTe$_{4}$ was cleaned in situ using argon ion milling to remove any residual oxide layer at the top interface. For the nonlocal device geometry, once the contacts were defined, TaIrTe$_{4}$ was then etched into strips by reactive ion etching with CF$_{4}$/O$_{2}$ employing a polymethyl methacrylate (PMMA) mask defined using electron beam lithography. We note the reversal in the order of etching and contacting the TaIrTe$_{4}$ flake was performed due to the slow etch rate of TaIrTe$_{4}$ to CF$_{4}$/O$_{2}$ that results in the cross-linking of the etch mask used. The final sample was spin-coated with PMMA to prevent the TaIrTe$_{4}$ strip from any further oxidation.
\\

\subsection*{Electrical transport measurements: } 

The various electrical measurements at different temperatures were carried out by varying the magnetic field in a superconducting solenoid cryostat (Cryogenic Limited). Based on the frequency dependence of the thermoelectric responses (see Supplementary Section X), we choose a driving frequency of 7.777 Hz for the ac current in the heater ensuring that the measurements are performed in the quasi-static thermal regime. The voltage was measured by standard lock-in techniques (SR830 and AMETEK 7270). The measured two-terminal resistances of the TaIrTe$_{4}$ strips are reported in Supplementary Section II.

\subsection*{Atomistic spin dynamics calculations}
Calculations of the magnon spectra and thermal conductivity were carried out using JAMS~\cite{barker_jams_2026}. The model was parametrised using exchange constants and anisotropy values derived from neutron scattering in Ref.~\citen{Calder_PhysRevB_102_024408_2020}. Our model includes dipole interactions and the value of anisotropy is adjusted to compensate for the dipole anisotropy. A supercell of $38^3$ primitive cells with periodic boundaries was used for the calculations. Magnon spectra were calculated from the spin-time Fourier transform of the spin-spin correlation functions. Magnon conductivity was calculated using the Kubo formalism. Full details are provided in Supplementary Section XIII.

\subsection*{Density functional theory (DFT)}
Calculations for CrPS$_4$ were carried out using the Vienna Ab initio Simulation Package (VASP)\cite{Kresse1996,Kresse1999}. Exchange and correlation were described within the generalized gradient approximation (GGA) using the Perdew--Burke--Ernzerhof (PBE) functional\cite{pbe_gga}. 
Van der Waals corrections were included via the DFT-D3 approach \cite{dft-d3}.
The tetrahedron method with Blöchl corrections was employed for the integration of the Brillouin zone \cite{tetra-vasp}. 
The crystal structure of CrPS$_4$ corresponds to the monoclinic space group C2, we considered the conventional unit cell with lattice parameters: $a=10.871$~\AA\, $b=7.254$~\AA\, $c        =6.140$~\AA\  and $\beta = 91.88^\circ$\ \cite{cps-ref}. 
To account for the magnetic ordering of the system, spin-polarized collinear calculation were carried out, considering an  A-type antiferromagnetic ground state.
The ionic positions were relaxed until the residual forces were below $10^{-3}$ eV/\AA.\\

Phonon spectra was calculated using the PHONOPY code via finite-differences method \cite{phonopy1,phonopy2}. 
PHONOPY was used to create 36 2x2x2 supercells of the conventional monoclinic unit cell (containing 192 atoms) with different atomic displacements. 
DFT calculations were then performed to obtain the forces acting on atoms.
For this goal, a plane-wave energy cutoff of 500 eV was used, with the total energy converged to $10^{-8}$ eV. BZ integrations were performed on a $4\times4\times4$ Monkhorst--Pack $k$-point mesh. 
Force constant calculations and post-processing were then carried out using PHONOPY. For the phonon density of states calculation a 41$^3$ mesh was used.
\section*{Data availability}
The data associated with this manuscript will be made available at DataverseNL at the moment of publication. 

\section*{Acknowledgements}

We thank S.H. Tirion and G.E.W. Bauer for fruitful discussions and comments on the manuscript. The authors acknowledge the technical support from J. G. Holstein, H. H. de Vries, H. Adema, F. H. van der Velde and A. Joshua. The authors thank Prof. M. Loi and M. Kot for providing technical support and access to their polarized Raman setup. K.S. and B.J.v.W. acknowledge the research program “Materials for the Quantum Age” (QuMat) for financial support. This program (registration number 024.005.006) is part of the Gravitation program financed by the Dutch Ministry of Education, Culture and Science (OCW). This work is additionally supported by the Zernike Institute for Advanced Materials (ZIAM) at University of Groningen, the Spinoza prize awarded to B.J. van Wees by the Nederlandse Organisatie voor Wetenschappelijk Onderzoek (NWO) in 2016, as well as by the European Research Council (ERC) under 2DMAGSPIN (Grant No. 101053054) and FERRERO (Grant No. 101170522). J.B. acknowledges funding from a Royal Society University Research Fellowship and the EPSRC Programme Grant CAMIE (EP/X027074/1). Spin dynamics calculations were performed on the Aire and Calder HPC system at the University of Leeds, UK. The first-principles calculations were carried out on the Dutch national e-infrastructure with the support of SURF Cooperative (EINF-17379) and on the Hábrók high performance computing cluster of the University of Groningen.
\section*{Author contributions}

K.S., S.A and B.J.v.W. conceived the experiment. K.S. and S.A. fabricated the devices and performed the electrical transport measurements. S.A. and K.S. analyzed the data with input and discussions on the experimental transport measurements from and B.J.v.W. C.A.C.S performed the polarized Raman measurements and performed the analysis and interpretation of the Raman data under the supervision of M.H.D.G. M.A. performed the scanning transmission electron microscopy. S.A. performed the DFT calculations with support of J.S. J.B. performed the atomistic spin dynamics. K.S., S.A. and J.B. wrote the manuscript with input from all authors. B.J.v.W. supervised the research.








\clearpage
\bibliography{Reference}

@article{Intro_2,
  title={Thermal Transport in Magnetic Materials: A Review},
  author={Li, Shuchen and Guo, Shucheng and Hoke, Thomas and Chen, Xi},
  journal={Materials Today Electronics},
  pages={100156},
  year={2025},
  publisher={Elsevier}
}

@article{Intro_1,
  title={Nanoscale thermal transport},
  author={Cahill, David G and Ford, Wayne K and Goodson, Kenneth E and Mahan, Gerald D and Majumdar, Arun and Maris, Humphrey J and Merlin, Roberto and Phillpot, Simon R},
  journal={Journal of Applied Physics},
  volume={93},
  number={2},
  pages={793--818},
  year={2003},
  publisher={American Institute of Physics}
}

@article{Intro_1b,
  title={{Nanoscale thermal transport. II. 2003--2012}},
  author={Cahill, David G and Braun, Paul V and Chen, Gang and Clarke, David R and Fan, Shanhui and Goodson, Kenneth E and Keblinski, Pawel and King, William P and Mahan, Gerald D and Majumdar, Arun and others},
  journal={Applied Physics Reviews},
  volume={1},
  number={1},
  year={2014},
  publisher={AIP Publishing}
}

@article{SpinCalo_1,
  title={Spin caloritronics},
  author={Bauer, Gerrit E W and Saitoh, Eiji and van Wees, Bart J},
  journal={Nature Materials},
  volume={11},
  number={5},
  pages={391--399},
  year={2012},
  publisher={Nature Publishing Group UK London}
}

@article{SpinCalo_2,
  title={Spin caloritronics},
  author={Boona, Stephen R and Myers, Roberto C and Heremans, Joseph P},
  journal={Energy \& Environmental Science},
  volume={7},
  number={3},
  pages={885--910},
  year={2014},
  publisher={Royal Society of Chemistry}
}

@article{ThermalValve_1,
  title={Room-temperature magnetic thermal switching by suppressing phonon-magnon scattering},
  author={Zhang, Fanghao and Patra, Lokanath and Chen, Yubi and Ouyang, Wenkai and Sarte, Paul M and Adajian, Shantal and Zuo, Xiangying and Yang, Runqing and Luo, Tengfei and Liao, Bolin},
  journal={Physical Review B},
  volume={109},
  number={18},
  pages={184411},
  year={2024},
  publisher={APS}
}

@article{ThermalValve_2,
  title={Emerging solid--state thermal switching materials},
  author={Jia, Junjun and Li, Shuchen and Chen, Xi and Shigesato, Yuzo},
  journal={Advanced Functional Materials},
  volume={34},
  number={42},
  pages={2406667},
  year={2024},
  publisher={Wiley Online Library}
}

@article{Thermal_Management,
  title={Two-dimensional materials for thermal management applications},
  author={Song, Houfu and Liu, Jiaman and Liu, Bilu and Wu, Junqiao and Cheng, Hui-Ming and Kang, Feiyu},
  journal={Joule},
  volume={2},
  number={3},
  pages={442--463},
  year={2018},
  publisher={Elsevier}
}

@article{AFM_1,
  title={Antiferromagnetic spintronics},
  author={Jungwirth, Tomas and Marti, X and Wadley, P and Wunderlich, J},
  journal={Nature Nanotechnology},
  volume={11},
  number={3},
  pages={231--241},
  year={2016},
  publisher={Nature Publishing Group}
}

@article{AFM_2,
  title={Antiferromagnetic spintronics},
  author={Baltz, Vincent and Manchon, Aurelien and Tsoi, M and Moriyama, Takahiro and Ono, T and Tserkovnyak, Y},
  journal={Reviews of Modern Physics},
  volume={90},
  number={1},
  pages={015005},
  year={2018},
  publisher={APS}
}

@article{SThM_SameHeaterDetector,
  title={Scanning thermal microscopy},
  author={Majumdar, Annu},
  journal={Annual review of materials science},
  volume={29},
  number={1},
  pages={505--585},
  year={1999},
  publisher={Annual Reviews 4139 El Camino Way, PO Box 10139, Palo Alto, CA 94303-0139, USA}
}

@techreport{RamanThermometry,
  title={Raman Thermometry of Microdevices: Comparing Methods to Minimize Error.},
  author={Beechem III, Thomas Edwin and Serrano, Justin Raymond},
  year={2011},
  institution={Sandia National Lab.(SNL-NM), Albuquerque, NM (United States)}
}

@article{SuspendedBridge,
  title={Reexamination of basal plane thermal conductivity of suspended graphene samples measured by electro-thermal micro-bridge methods},
  author={Jo, Insun and Pettes, Michael T and Lindsay, Lucas and Ou, Eric and Weathers, Annie and Moore, Arden L and Yao, Zhen and Shi, Li},
  journal={AIP Advances},
  volume={5},
  number={5},
  year={2015},
  publisher={AIP Publishing}
}

@article{ThermalHall_1,
  title={{Role of magnetic ions in the thermal Hall effect of the paramagnetic insulator TmVO$_4$}},
  author={Vallipuram, Ashvini and Chen, Lu and Campillo, Emma and Mezidi, Manel and Grissonnanche, Ga{\"e}l and Boulanger, Marie-Eve and Lefran{\c{c}}ois, {\'E}tienne and Zic, Mark P and Li, Yuntian and Fisher, Ian R and others},
  journal={Physical Review B},
  volume={110},
  number={4},
  pages={045144},
  year={2024},
  publisher={APS}
}

@article{SpinSeebeck,
  title={{Theory of the spin Seebeck effect}},
  author={Adachi, Hiroto and Uchida, Ken-ichi and Saitoh, Eiji and Maekawa, Sadamichi},
  journal={Reports on Progress in Physics},
  volume={76},
  number={3},
  pages={036501},
  year={2013},
  publisher={IOP Publishing}
}

@article{MagnonThermalTransport,
  title={Effect of magnons on thermal transport in insulators},
  author={Rives, J E and Dixon, G S and Walton, D},
  journal={Journal of Applied Physics},
  volume={40},
  number={3},
  pages={1555--1556},
  year={1969},
  publisher={American Institute of Physics}
}

@article{CrI3_ThermalHall,
  title={{Thermal Hall effect in the van der Waals ferromagnet CrI$_3$}},
  author={Xu, Chunqiang and Zhang, Heda and Carnahan, Caitlin and Zhang, Pengpeng and Xiao, Di and Ke, Xianglin},
  journal={Physical Review B},
  volume={109},
  number={9},
  pages={094415},
  year={2024},
  publisher={APS}
}

@article{VI3_ThermalHall,
  title={{Anomalous thermal Hall effect in an insulating van der Waals magnet}},
  author={Zhang, Heda and Xu, Chunqiang and Carnahan, Caitlin and Sretenovic, Milos and Suri, Nishchay and Xiao, Di and Ke, Xianglin},
  journal={Physical Review Letters},
  volume={127},
  number={24},
  pages={247202},
  year={2021},
  publisher={APS}
}

@article{Kappa_CC3,
  title={Giant thermal magnetoconductivity in {CrCl$_3$} and a general model for spin-phonon scattering},
  author={Pocs, Christopher A and Leahy, Ian A and Zheng, Hao and Cao, Gang and Choi, Eun-Sang and Do, S-H and Choi, Kwang-Yong and Normand, B and Lee, Minhyea},
  journal={Physical Review Research},
  volume={2},
  number={1},
  pages={013059},
  year={2020},
  publisher={APS}
}

@article{Callaway1,
  title={Model for lattice thermal conductivity at low temperatures},
  author={Callaway, Joseph},
  journal={Physical Review},
  volume={113},
  number={4},
  pages={1046},
  year={1959},
  publisher={APS}
}

@article{Callaway2,
  title={Foundations of the Callaway theory of thermal conductivity},
  author={Nettleton, RE},
  journal={Physical Review},
  volume={132},
  number={5},
  pages={2032},
  year={1963},
  publisher={APS}
}

@article{Mag_Ph1,
  title={{Magnetic field enhancement of heat transport in the 2D Heisenberg antiferromagnet K$_2$V$_3$O$_8$}},
  author={Sales, B C and Lumsden, M D and Nagler, S E and Mandrus, D and Jin, R},
  journal={Physical Review Letters},
  volume={88},
  number={9},
  pages={095901},
  year={2002},
  publisher={APS}
}

@article{Mag_Ph2,
  title={Spin-Phonon Scattering-Induced Low Thermal Conductivity in a {van der Waals} Layered Ferromagnet {Cr$_2$Si$_2$Te$_6$}},
  author={Yang, Kunya and Wu, Hong and Li, Zefang and Ran, Chen and Wang, Xiao and Zhu, Fengfeng and Gong, Xiangnan and Liu, Yan and Wang, Guiwen and Zhang, Long and others},
  journal={Advanced Functional Materials},
  volume={33},
  number={37},
  pages={2302191},
  year={2023},
  publisher={Wiley Online Library}
}

@article{Uchida_Trans_Seebeck,
  title={Thermoelectrics: From longitudinal to transverse},
  author={Uchida, Ken-ichi and Heremans, Joseph P},
  journal={Joule},
  volume={6},
  number={10},
  pages={2240--2245},
  year={2022},
  publisher={Elsevier}
}

@article{KMgBi,
  title={Synergizing a Large Ordinary {Nernst} Effect and Axis-Dependent Conduction Polarity in Flat Band {KMgBi} Crystals},
  author={Ochs, Andrew M and Fecher, Gerhard H and He, Bin and Schnelle, Walter and Felser, Claudia and Heremans, Joseph P and Goldberger, Joshua E},
  journal={Advanced Materials},
  volume={36},
  number={2},
  pages={2308151},
  year={2024},
  publisher={Wiley Online Library}
}

@article{NaSnAs,
  title={Computationally guided discovery of axis-dependent conduction polarity in {NaSnAs} crystals},
  author={Ochs, Andrew M and Gorai, Prashun and Wang, Yaxian and Scudder, Michael R and Koster, Karl and Moore, Curtis E and Stevanovic, Vladan and Heremans, Joseph P and Windl, Wolfgang and Toberer, Eric S and others},
  journal={Chemistry of Materials},
  volume={33},
  number={3},
  pages={946--951},
  year={2021},
  publisher={ACS Publications}
}

@article{LaPt2B,
  title={{Large transverse thermoelectric effect induced by the mixed-dimensionality of Fermi surfaces}},
  author={Manako, Hikari and Ohsumi, Shoya and Sato, Yoshiki J and Okazaki, R and Aoki, D},
  journal={Nature Communications},
  volume={15},
  number={1},
  pages={3907},
  year={2024},
  publisher={Nature Publishing Group UK London}
}

@article{WSi2,
  title={Transverse Thermoelectric Conversion in the Mixed-Dimensional Semimetal {WSi$_{2}$}},
  author={Ohsumi, Shoya and Sato, Yoshiki J and Okazaki, Ryuji},
  journal={PRX Energy},
  volume={3},
  number={4},
  pages={043007},
  year={2024},
  publisher={APS}
}

@article{TaIrTe4_Gonio,
  title={{p x n-Type} Transverse Thermoelectrics in a {Type-II Weyl} Semimetal {TaIrTe$_{4}$}},
  author={Mutch, Josh and Chen, Wei-Chih and Huang, Cheng-Yi and Malinowski, Paul and Chen, Cheng-Chien and Chu, Jiun-Haw},
  journal={arXiv preprint arXiv:2207.13687},
  year={2022) (accessed February 6, 2026}
}

@article{TaIrTe4_Gonio_2,
  title={Large Transverse Thermoelectric Effect in {Weyl} Semimetal {TaIrTe$_4$} Engineered for Photodetection},
  author={Blevins, Morgan G and Ji, Xianglin and Santamaria-Garcia, Vivian J and Mukherjee, Abhishek and Nguyen, Thanh and Li, Mingda and Boriskina, Svetlana V},
  journal={arXiv preprint arXiv:2602.14959},
  year={2026) (accessed March 2, 2026}
}

@article{shao2025abnormal,
  title={Abnormal thermal conductivity in thin antiferromagnet {CrPS$_4$} nanoribbons: Effect of elastic stiffening and magnon--phonon coupling},
  author={Shao, Shuai and He, Runzhi and Chen, Wenkang and Zeng, Xiaoliang and Xu, Xiangfan},
  journal={APL Materials},
  volume={13},
  number={9},
  year={2025},
  publisher={AIP Publishing}
}

@article{NL_HighField_CPS,
  title={{Magnon spin transport in the van der Waals antiferromagnet CrPS$_{4}$ for noncollinear and collinear magnetization}},
  author={De Wal, Dennis K and Zohaib, Muhammad and van Wees, Bart J},
  journal={Physical Review B},
  volume={110},
  number={17},
  pages={174440},
  year={2024},
  publisher={APS}
}

@article{HallBar_CPS,
  title={{Spin Seebeck in the weakly exchange-coupled van der Waals antiferromagnet across the spin-flip transition}},
  author={He, Xue and Ding, Shilei and Giil, Hans Gl{\o}ckner and Wang, Jicheng and Bhukta, Mona and Wu, Mingxing and Shi, Wen and Lin, Zhongchong and Liang, Zhongyu and Yang, Jinbo and others},
  journal={Nature Communications},
  volume={16},
  number={1},
  pages={3037},
  year={2025},
  publisher={Nature Publishing Group UK London}
}

@article{SQUID_CPS,
  title={{Magnetic structure and metamagnetic transitions in the van der Waals antiferromagnet CrPS$_{4}$}},
  author={Peng, Yuxuan and Ding, Shilei and Cheng, Man and Hu, Qifeng and Yang, Jie and Wang, Fanggui and Xue, Mingzhu and Liu, Zhou and Lin, Zhongchong and Avdeev, Maxim and others},
  journal={Advanced Materials},
  volume={32},
  number={28},
  pages={2001200},
  year={2020},
  publisher={Wiley Online Library}
}

@article{ACS_Nano,
  title={Toward Two-Dimensional {van der Waals} Magnon Transport Devices: {WTe$_2$} Electrodes for Efficient Magnon Spin Injection and Detection},
  author={Sundararajan, Krishnaraajan and de Wal, Dennis K and Alvarruiz, Sergio and Cordero-Silis, C{\'e}dric A and Ahmadi, Majid and Guimar{\~a}es, Marcos H D and van Wees, Bart J},
  journal={ACS Nano},
  volume={19},
  number={44},
  pages={38716--38723},
  year={2025},
  publisher={ACS Publications}
}

@article{1_Zomer,
  title={Fast pick up technique for high quality heterostructures of bilayer graphene and hexagonal boron nitride},
  author={Zomer, P J and Guimar{\~a}es, M H D and Brant, J C and Tombros, N and van Wees, B J},
  journal={Applied Physics Letters},
  volume={105},
  number={1},
  year={2014},
  publisher={AIP Publishing}
}

@article{Mag_MPS,
  title={{Magnon transport in quasi-two-dimensional van der Waals antiferromagnets}},
  author={Xing, Wenyu and Qiu, Luyi and Wang, Xirui and Yao, Yunyan and Ma, Yang and Cai, Ranran and Jia, Shuang and Xie, XC and Han, Wei},
  journal={Physical Review X},
  volume={9},
  number={1},
  pages={011026},
  year={2019},
  publisher={APS}
}

@article{DKW_1,
  title={{Long-distance magnon transport in the van der Waals antiferromagnet CrPS$_4$}},
  author={De Wal, Dennis K and Iwens, Arnaud and Liu, Tian and Tang, Ping and Bauer, Gerrit E W and van Wees, Bart J},
  journal={Physical Review B},
  volume={107},
  number={18},
  pages={L180403},
  year={2023},
  publisher={APS}
}

@article{nakagawa,
  title={Observation of room temperature intrinsic nonlinear thermoelectric effects in low-dimensional semimetals},
  author={Nakagawa, Kurea and Sundararajan, Krishnaraajan and Cordero-Silis, C{\'e}dric A and van Wees, Bart J and Guimar{\~a}es, Marcos H D},
  journal={arXiv preprint arXiv:2607.21808},
  year={2026) (accessed August 22, 2026}
}

@article{arxiv_Aniso,
   title={{Anisotropic magnon spin transport in CrPS$_4$}},
  author={Sundararajan, Krishnaraajan and Zohaib, Muhammad and Kreminska, Yuliia and Tirion, Sytze H and van Wees, Bart J},
  journal={arXiv preprint arXiv:2608.10802},
  year={2026) (accessed August 22, 2026}
}

@article{phonopy1,
  author  = {Togo, Atsushi and Chaput, Laurent and Tadano, Terumasa and Tanaka, Isao},
  title   = {Implementation strategies in phonopy and phono3py},
  journal = {Journal  of Physics: Condensed Matter},
  volume  = {35},
  number  = {35},
  pages   = {353001},
  year    = {2023},
  doi     = {10.1088/1361-648X/acd831}
}

@article{phonopy2,
  author  = {Togo, Atsushi},
  title   = {First-principles phonon calculations with phonopy and phono3py},
  journal = {Journal of the Physical Society of Japan},
  volume  = {92},
  number  = {1},
  pages   = {012001},
  year    = {2023},
  doi     = {10.7566/JPSJ.92.012001}
}

@article{dft-d3,
    author = {Grimme, Stefan and Antony, Jens and Ehrlich, Stephan and Krieg, Helge},
    title = {{A consistent and accurate ab initio parametrization of density functional dispersion correction (DFT-D) for the 94 elements H-Pu}},
    journal = {The Journal of Chemical Physics},
    volume = {132},
    number = {15},
    pages = {154104},
    year = {2010},
    month = {04},
    issn = {0021-9606},
    doi = {10.1063/1.3382344},
    url = {https://doi.org/10.1063/1.3382344},
}

@article{tetra-vasp,
  title = {{Improved tetrahedron method for Brillouin-zone integrations}},
  author = {Bl\"ochl, Peter E. and Jepsen, O. and Andersen, O. K.},
  journal={Physical Review B},
  volume = {49},
  issue = {23},
  pages = {16223--16233},
  numpages = {0},
  year = {1994},
  month = {Jun},
  publisher = {American Physical Society},
  doi = {10.1103/PhysRevB.49.16223},
  url = {https://link.aps.org/doi/10.1103/PhysRevB.49.16223}
}

@article{cps-ref,
author = "Diehl, R. and Carpentier, C.-D.",
title = "{The crystal structure of chromium thiophosphate, CrPS$_{4}$}",
journal = "Acta Crystallographica Section B",
year = "1977",
volume = "33",
number = "5",
pages = "1399--1404",
month = "May",
doi = {10.1107/S0567740877006165},
url = {https://doi.org/10.1107/S0567740877006165},
}

@article{Kresse1996,
  title={Efficient iterative schemes for ab initio total-energy calculations using a plane-wave basis set},
  author={Kresse, Georg and Furthm{\"u}ller, J{\"u}rgen},
  journal={Physical Review B},
  volume={54},
  number={16},
  pages={11169},
  year={1996},
  publisher={APS}
}

@article{Kresse1999,
  title={From ultrasoft pseudopotentials to the projector augmented-wave method},
  author={Kresse, Georg and Joubert, Daniel},
  journal={Physical Review B},
  volume={59},
  number={3},
  pages={1758},
  year={1999},
  publisher={APS}
}

@article{pbe_gga,
  title={Generalized gradient approximation made simple},
  author={Perdew, John P and Burke, Kieron and Ernzerhof, Matthias},
  journal={Physical Review Letters},
  volume={77},
  number={18},
  pages={3865},
  year={1996},
  publisher={American Physical Society}
}

@article{Liu2023_raman,
  title = {Field-free switching of perpendicular magnetization at room temperature using out-of-plane spins from {TaIrTe$_4$}},
  volume = {6},
  ISSN = {2520-1131},
  number = {10},
  journal = {Nature Electronics},
  publisher = {Springer Science and Business Media LLC},
  author = {Liu,  Yakun and Shi,  Guoyi and Kumar,  Dushyant and Kim,  Taeheon and Shi,  Shuyuan and Yang,  Dongsheng and Zhang,  Jiantian and Zhang,  Chenhui and Wang,  Fei and Yang,  Shuhan and Pu,  Yuchen and Yu,  Peng and Cai,  Kaiming and Yang,  Hyunsoo},
  year = {2023},
  month = oct,
  pages = {732–738}
}

@article{Liu2018_raman,
  title = {Raman Signatures of Broken Inversion Symmetry and In‐Plane Anisotropy in {Type‐II} Weyl Semimetal Candidate {TaIrTe$_4$}},
  volume = {30},
  ISSN = {1521-4095},
  number = {25},
  journal = {Advanced Materials},
  publisher = {Wiley},
  author = {Liu,  Yinan and Gu,  Qiangqiang and Peng,  Yu and Qi,  Shaomian and Zhang,  Na and Zhang,  Yinong and Ma,  Xiumei and Zhu,  Rui and Tong,  Lianming and Feng,  Ji and Liu,  Zheng and Chen,  Jian‐Hao},
  year = {2018},
  month = may 
}

@article{luo2026flat,
  title={Flat band induced quasi-one-dimensional magnon transport in a two-dimensional spin lattice},
  author={Luo, Bingcheng and Chen, Mantang and Wang, Ziqian and Xia, Zi’an and Zhai, Yuxin and Jia, Linhao and Cui, Hang and Song, Yuanjun and Xiong, Qihua and Xiao, Jiang and others},
  journal={Nature Communications},
  volume={17},
  number={1},
  pages={4292},
  year={2026},
  publisher={Nature Publishing Group UK London}
}

@article{qi2023giant,
  title={{Giant electrically tunable magnon transport anisotropy in a van der Waals antiferromagnetic insulator}},
  author={Qi, Shaomian and Chen, Di and Chen, Kangyao and Liu, Jianqiao and Chen, Guangyi and Luo, Bingcheng and Cui, Hang and Jia, Linhao and Li, Jiankun and Huang, Miaoling and others},
  journal={Nature Communications},
  volume={14},
  number={1},
  pages={2526},
  year={2023},
  publisher={Nature Publishing Group UK London}
}

@article{he2019fermi,
  title={{The Fermi surface geometrical origin of axis-dependent conduction polarity in layered materials}},
  author={He, Bin and Wang, Yaxian and Arguilla, Maxx Q and Cultrara, Nicholas D and Scudder, Michael R and Goldberger, Joshua E and Windl, Wolfgang and Heremans, Joseph P},
  journal={Nature Materials},
  volume={18},
  number={6},
  pages={568--572},
  year={2019},
  publisher={Nature Publishing Group UK London}
}

@article{Buttiker_TenFold,
  title={Onsager relations in coupled electric, thermoelectric, and spin transport: The tenfold way},
  author={Jacquod, Philippe and Whitney, Robert S and Meair, Jonathan and B{\"u}ttiker, Markus},
  journal={Physical Review B},
  volume={86},
  number={15},
  pages={155118},
  year={2012},
  publisher={APS}
}

@article{Onsager_A,
  title={{Reciprocal relations in irreversible processes. I.}},
  author={Onsager, Lars},
  journal={Physical Review},
  volume={37},
  number={4},
  pages={405},
  year={1931},
  publisher={APS}
}

@article{Application_Onsager,
  title={{The application of Onsager's reciprocal relations to thermoelectric, thermomagnetic, and galvanomagnetic effects}},
  author={Callen, Herbert B},
  journal={Physical Review},
  volume={73},
  number={11},
  pages={1349},
  year={1948},
  publisher={APS}
}

@article{Cahill_3omega,
  title={Thermal conductivity of amorphous solids above the plateau},
  author={Cahill, David G and Pohl, Robert O},
  journal={Physical Review B},
  volume={35},
  number={8},
  pages={4067},
  year={1987},
  publisher={APS}
}

@article{Tutorial_3omega,
  title={A tutorial on the {3$\omega$} method for measuring thermal conductivity from macro to micro dimensions},
  author={Yang, Xiao and Zhao, Haibo and Wang, Chunyang and Shen, Yanan and Song, Zewen and Chen, Haisheng and Zhang, Ting and Zheng, Xinghua},
  journal={Journal of Applied Physics},
  volume={138},
  number={21},
  year={2025},
  publisher={AIP Publishing}
}

@article{bud2021magnetic,
  title={{Magnetic order in the van der Waals antiferromagnet CrPS$_4$: Anisotropic H-T phase diagrams and effects of pressure}},
  author={Bud'ko, Sergey L and Gati, Elena and Slade, Tyler J and Canfield, Paul C},
  journal={Physical Review B},
  volume={103},
  number={22},
  pages={224407},
  year={2021},
  publisher={APS}
}

@article{cui2023chirality,
  title={Chirality selective magnon-phonon hybridization and magnon-induced chiral phonons in a layered zigzag antiferromagnet},
  author={Cui, Jun and Bostr{\"o}m, Emil Vi{\~n}as and Ozerov, Mykhaylo and Wu, Fangliang and Jiang, Qianni and Chu, Jiun-Haw and Li, Changcun and Liu, Fucai and Xu, Xiaodong and Rubio, Angel and others},
  journal={Nature Communications},
  volume={14},
  number={1},
  pages={3396},
  year={2023},
  publisher={Nature Publishing Group UK London}
}

@article{fang2025efficient,
  title={Efficient method for calculating magnon-phonon coupling from first principles},
  author={Fang, Wuzhang and Simoni, Jacopo and Ping, Yuan},
  journal={Physical Review B},
  volume={111},
  number={10},
  pages={104431},
  year={2025},
  publisher={APS}
}

@article{jensen1975spin,
  title={{Spin waves in terbium. II. Magnon-phonon interaction}},
  author={Jensen, J and Houmann, J G},
  journal={Physical Review B},
  volume={12},
  number={1},
  pages={320},
  year={1975},
  publisher={APS}
}

@article{Calder_PhysRevB_102_024408_2020,
  author    = {Calder, S. and Haglund, A. V. and Liu, Y. and Pajerowski, D. M. and Cao, H. B. and Williams, T. J. and Garlea, V. O. and Mandrus, D.},
  title     = {{Magnetic} structure and exchange interactions in the layered semiconductor {CrPS$_4$}},
  journal   = {Phys. Rev. B},
  number    = {2},
  volume    = {102},
  pages     = {024408},
  year      = {2020},
  month     = {7},
  publisher = {American Physical Society (APS)},
  doi       = {10.1103/physrevb.102.024408},
  url       = {https://dx.doi.org/10.1103/physrevb.102.024408}
}

@misc{barker_jams_2026,
  author       = {Barker, Joseph},
  title        = {{JAMS}: Joe's Awesome Magnetic Simulator},
  year         = {2026},
  publisher    = {Zenodo},
  doi          = {10.5281/zenodo.19330399},
  url          = {https://doi.org/10.5281/zenodo.19330399},
  note         = {{DOI}: \url{https://doi.org/10.5281/zenodo.19330399};
                  source code: \url{https://github.com/stonerlab/jams}}
}

@article{Kikkawa_PhysRevLett_117_207203_2016,
  author    = {Kikkawa, Takashi and Shen, Ka and Flebus, Benedetta and Duine, Rembert A. and Uchida, Ken-ichi and Qiu, Zhiyong and Bauer, Gerrit E. W. and Saitoh, Eiji},
  title     = {{Magnon} {Polarons} in the {Spin} {Seebeck} {Effect}},
  journal   = {Phys. Rev. Lett.},
  number    = {20},
  volume    = {117},
  pages     = {207203},
  year      = {2016},
  month     = {11},
  publisher = {American Physical Society (APS)},
  doi       = {10.1103/physrevlett.117.207203},
  url       = {https://dx.doi.org/10.1103/physrevlett.117.207203}
}

@article{Ramos_NatCommun_10_5162_2019,
  author    = {Ramos, R. and Hioki, T. and Hashimoto, Y. and Kikkawa, T. and Frey, P. and Kreil, A. J. E. and Vasyuchka, V. I. and Serga, A. A. and Hillebrands, B. and Saitoh, E.},
  title     = {{Room} temperature and low-field resonant enhancement of spin {Seebeck} effect in partially compensated magnets},
  journal   = {Nat. Commun.},
  number    = {1},
  volume    = {10},
  pages     = {5162},
  year      = {2019},
  month     = {11},
  publisher = {Springer Science and Business Media LLC},
  doi       = {10.1038/s41467-019-13121-5},
  url       = {https://dx.doi.org/10.1038/s41467-019-13121-5}
}

@article{Flebus_PhysRevB_95_144420_2017,
  author    = {Flebus, Benedetta and Shen, Ka and Kikkawa, Takashi and Uchida, Ken-ichi and Qiu, Zhiyong and Saitoh, Eiji and Duine, Rembert A. and Bauer, Gerrit E. W.},
  title     = {{Magnon}-polaron transport in magnetic insulators},
  journal   = {Phys. Rev. B},
  number    = {14},
  volume    = {95},
  pages     = {144420},
  year      = {2017},
  month     = {4},
  publisher = {American Physical Society (APS)},
  doi       = {10.1103/physrevb.95.144420},
  url       = {https://dx.doi.org/10.1103/physrevb.95.144420}
}

@misc{Regout_2605_07416_2026,
  author    = {Regout, Mexx. E. Y. and Dup{\'e}, Bertrand and Verstraete, Matthieu J.},
  title     = {{Effective} {Gilbert} damping in the stochastic {Landau}-{Lifshitz}-{Gilbert} equation},
  journal   = {arXiv.2605.07416v1 [cond-mat.mes-hall]},
  year      = {2026},
  month     = {5},
  eprint    = {2605.07416},
  archiveprefix = {arXiv},
  primaryclass = {cond-mat.mes-hall},
  url       = {https://arxiv.org/abs/2605.07416v1}
}

@article{hentrich2018unusual,
  title={Unusual phonon heat transport in {$\alpha$-RuCl$_3$}: {Strong} spin-phonon scattering and field-induced spin gap},
  author={Hentrich, Richard and Wolter, Anja UB and Zotos, Xenophon and Brenig, Wolfram and Nowak, Domenic and Isaeva, Anna and Doert, Thomas and Banerjee, Arnab and Lampen-Kelley, Paula and Mandrus, David G and others},
  journal={Physical review letters},
  volume={120},
  number={11},
  pages={117204},
  year={2018},
  publisher={APS}
}
\clearpage

\foreach \p in {1,...,35}{%
  \clearpage
  \includepdf[
    pages={\p},
    pagecommand={},
    turn=false
  ]{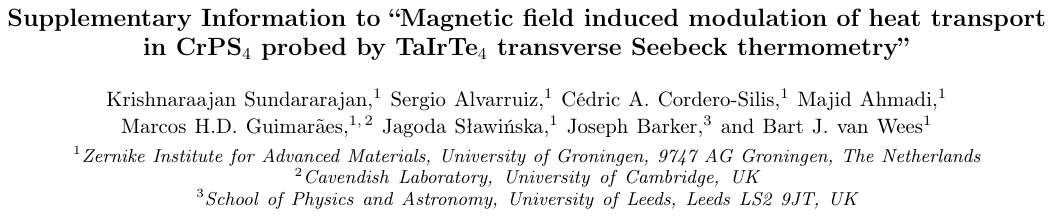}%
}

\end{document}